\documentclass[aps,onecolumn]{revtex4}
\usepackage{amssymb,epsf}
\usepackage{latexsym}

\begin{document}

\title{Thermodynamics of third order Lovelock adS black holes \\
in the presence of Born-Infeld type nonlinear electrodynamics}
\author{S. H. Hendi$^{1,2}$\footnote{email address: hendi@shirazu.ac.ir} and A. Dehghani$^1$}
\affiliation{$^1$ Physics Department and Biruni Observatory, College of Sciences, Shiraz
University, Shiraz 71454, Iran \\
$^2$ Research Institute for Astrophysics and Astronomy of Maragha (RIAAM),
P.O. Box 55134-441, Maragha, Iran}

\begin{abstract}
In this paper, we obtain topological black hole solutions of third
order Lovelock gravity couple with two classes of Born-Infeld type
nonlinear electrodynamics with anti-de Sitter asymptotic
structure. We investigate geometric and thermodynamics properties
of the solutions and obtain conserved quantities of the black
holes. We examine the first law of thermodynamics and find that
the conserved and thermodynamic quantities of the black hole
solutions satisfy the first law of thermodynamics. Finally, we
calculate the heat capacity and determinant of Hessian matrix to
evaluate thermal stability in both canonical and grand canonical
ensembles. Moreover, we consider extended phase space
thermodynamics to obtain generalized first law of thermodynamics
as well as extended Smarr formula.
\end{abstract}

\maketitle

\section{Introduction}

Regarding string theory and brane world cosmology, it has been shown that
spacetimes possess more than four dimensions. Taking into account higher
dimensional spacetimes, we know that the general conserved symmetric tensor
that depends on the metric and its derivatives up to second order is not the
Einstein tensor.

One of the natural generalization of Einstein theory in higher dimensional
spacetimes, in which contains most of Einstein assumptions, is Lovelock
gravity \cite{Lovelock}. Lovelock gravity field equation contains metric
derivatives no higher second order and therefore its quantization is free of
ghost \cite{GhostFree}. Since the higher curvature terms of Lovelock
Lagrangian appear in the low-energy limit of string theory, black hole
solutions of Lovelock gravity have been attracting renewed interest. The
action of Lovelock gravity in a compressed form can be written as
\begin{equation}
I_{G}=\int d^{d}x\sqrt{-g}\sum\limits_{k=0}^{d/2}\alpha _{k}\pounds _{k},
\label{LovelockAction}
\end{equation}
where $\alpha_{k}$'s are arbitrary constants and $\pounds _{k}$'s are the
Euler densities of the $2k$-dimensional manifolds with the following
explicit form
\begin{equation}
\pounds _{k}=\delta _{\rho _{1}\sigma _{1}...\rho _{k}\sigma _{k}}^{\mu
_{1}\nu _{1}...\mu _{k}\nu _{k}}R_{\mu _{1}\nu _{1}}^{\rho _{1}\sigma
_{1}}...R_{\mu _{k}\nu _{k}}^{\rho _{k}\sigma _{k}}.  \label{Lk}
\end{equation}%
In Eq. (\ref{Lk}), $\delta _{\rho _{1}\sigma _{1}...\rho _{k}\sigma
_{k}}^{\mu _{1}\nu _{1}...\mu _{k}\nu _{k}}$ and $R_{\mu \nu }^{\rho \sigma
} $ are, respectively, the generalized totally anti-symmetric Kronecker
delta and the Riemann tensor.

The objective of this paper is to find topological (asymptotically
adS) black hole solutions of third order Lovelock gravity in the
presence of two classes of Born-Infeld (BI) type nonlinear
electrodynamics (NED). Some black object solutions of third order
Lovelock theory coupled with NED have been studied before
\cite{TOLNLED}. Recently, one of us considered BI type Lagrangians
to obtain the black hole solutions \cite{HendiJHEP,HendiAnn}. The
Lagrangians of exponential and logarithmic forms of BI type
theories may be defined as
\begin{equation}
\mathcal{L}(\mathcal{F})=\left\{
\begin{array}{ll}
\beta ^{2}\left( \exp (-\frac{\mathcal{F}}{\beta ^{2}})-1\right) , & \;~%
\text{{ENED}} \\
-8\beta ^{2}\ln \left( 1+\frac{\mathcal{F}}{8\beta ^{2}}\right) , & \;\text{~%
{LNED}}%
\end{array}%
\right. ,  \label{LF}
\end{equation}%
where $\beta $ is called the nonlinearity parameter, the Maxwell
invariant is $\mathcal{F}=F_{\mu \nu }F^{\mu \nu }$ in which
$F_{\mu \nu }=\partial _{\mu }A_{\nu }-\partial _{\nu }A_{\mu }$
is the Faraday tensor and $A_{\mu } $ is the gauge potential.
Although BI type models was introduced with various motivations,
the important motivation of considering the BI type NED theories
comes from the fact that these theories may be originated if one
regards the loop corrections \cite{Fradkin85}. Taking into account
the coupling of BI type theories with Einstein and Gauss-Bonnet
gravity, it was shown that although there are some analogies
between the BI type theories,
there exist some differences between them as well \cite%
{HendiJHEP,HendiAnn,HendiNon,HendiMahmudi}. Now, we would like to obtain new
topological black hole solutions of the mentioned models of BI type theories
coupled with third order Lovelock gravity and investigate their geometric
and thermodynamic properties.

The outline of the paper is as follows. We present the topological black
hole solutions in Sec. \ref{Topol}. Sec. \ref{Thermodynamics}, is devoted to
investigate conserved and thermodynamic quantities of topological black
holes. We also analyze the thermodynamic stability of the solutions in the
canonical and grand canonical ensembles. We finish our paper with some
conclusions.

\section{Topological adS black holes in third order Lovelock gravity \label%
{Topol}}

The gravitational and electromagnetic field equations of third
order Lovelock gravity in the presence of NED may be written as
\begin{equation}
G_{\mu \nu }+\Lambda g_{\mu \nu }+\alpha _{2}\mathcal{G}_{\mu \nu }+\alpha
_{3}\mathcal{H}_{\mu \nu }=\frac{1}{2}g_{\mu \nu }\mathcal{L}(\mathcal{F}%
)-2F_{\mu \lambda }F_{\nu }^{\;\lambda }\mathcal{L}_{\mathcal{F}},
\label{Geq}
\end{equation}
\begin{equation}
\partial _{\mu }\left( \sqrt{-g}\mathcal{L}_{\mathcal{F}}F^{\mu \nu }\right)
=0,  \label{BIeq}
\end{equation}%
where $G_{\mu \nu }$ is the Einstein tensor, $\mathcal{G}_{\mu \nu }$ and $%
\mathcal{H}_{\mu \nu }$ are, respectively, the second and third orders
Lovelock tensor given as
\begin{equation}
\mathcal{G}_{\mu \nu } =2(R_{\mu \sigma \kappa \tau }R_{\nu }^{\phantom{\nu}%
\sigma \kappa \tau }-2R_{\mu \rho \nu \sigma }R^{\rho \sigma }-2R_{\mu
\sigma }R_{\phantom{\sigma}\nu }^{\sigma }+RR_{\mu \nu })-\frac{1}{2}g_{\mu
\nu }\mathcal{L}_{2},  \label{Love2}
\end{equation}
\begin{eqnarray}
\mathcal{H}_{\mu \nu } &=&-3[4R^{\tau \rho \sigma \kappa }R_{\sigma \kappa
\lambda \rho }R_{\phantom{\lambda }{\nu \tau \mu}}^{\lambda }-8R_{%
\phantom{\tau \rho}{\lambda \sigma}}^{\tau \rho }R_{\phantom{\sigma
\kappa}{\tau \mu}}^{\sigma \kappa }R_{\phantom{\lambda }{\nu \rho \kappa}%
}^{\lambda }+2R_{\nu }^{\phantom{\nu}{\tau \sigma \kappa}}R_{\sigma \kappa
\lambda \rho }R_{\phantom{\lambda \rho}{\tau \mu}}^{\lambda \rho }  \nonumber
\\
&&-R^{\tau \rho \sigma \kappa }R_{\sigma \kappa \tau \rho }R_{\nu \mu }+8R_{%
\phantom{\tau}{\nu \sigma \rho}}^{\tau }R_{\phantom{\sigma \kappa}{\tau \mu}%
}^{\sigma \kappa }R_{\phantom{\rho}\kappa }^{\rho }+8R_{\phantom
{\sigma}{\nu \tau \kappa}}^{\sigma }R_{\phantom {\tau \rho}{\sigma \mu}%
}^{\tau \rho }R_{\phantom{\kappa}{\rho}}^{\kappa }  \nonumber \\
&&+4R_{\nu }^{\phantom{\nu}{\tau \sigma \kappa}}R_{\sigma \kappa \mu \rho
}R_{\phantom{\rho}{\tau}}^{\rho }-4R_{\nu }^{\phantom{\nu}{\tau \sigma
\kappa }}R_{\sigma \kappa \tau \rho }R_{\phantom{\rho}{\mu}}^{\rho
}+4R^{\tau \rho \sigma \kappa }R_{\sigma \kappa \tau \mu }R_{\nu \rho
}+2RR_{\nu }^{\phantom{\nu}{\kappa \tau \rho}}R_{\tau \rho \kappa \mu }
\nonumber \\
&&+8R_{\phantom{\tau}{\nu \mu \rho }}^{\tau }R_{\phantom{\rho}{\sigma}%
}^{\rho }R_{\phantom{\sigma}{\tau}}^{\sigma }-8R_{\phantom{\sigma}{\nu \tau
\rho }}^{\sigma }R_{\phantom{\tau}{\sigma}}^{\tau }R_{\mu }^{\rho }-8R_{%
\phantom{\tau }{\sigma \mu}}^{\tau \rho }R_{\phantom{\sigma}{\tau }}^{\sigma
}R_{\nu \rho }-4RR_{\phantom{\tau}{\nu \mu \rho }}^{\tau }R_{\phantom{\rho}%
\tau }^{\rho }  \nonumber \\
&&+4R^{\tau \rho }R_{\rho \tau }R_{\nu \mu }-8R_{\phantom{\tau}{\nu}}^{\tau
}R_{\tau \rho }R_{\phantom{\rho}{\mu}}^{\rho }+4RR_{\nu \rho }R_{%
\phantom{\rho}{\mu }}^{\rho }-R^{2}R_{\nu \mu }]-\frac{1}{2}g_{\mu \nu }%
\mathcal{L}_{3}  \label{Love3}
\end{eqnarray}

In the recent equations, $\mathcal{L}_{2}$ and $\mathcal{L}_{3}$ denote the
Gauss-Bonnet Lagrangian and the third order Lovelock term, given as
\begin{equation}
\mathcal{L}_{2}=R_{\mu \nu \gamma \delta }R^{\mu \nu \gamma \delta }-4R_{\mu
\nu }R^{\mu \nu }+R^{2},  \label{L2}
\end{equation}%
\begin{eqnarray}
\mathcal{L}_{3} &=&2R^{\mu \nu \sigma \kappa }R_{\sigma \kappa \rho \tau }R_{%
\phantom{\rho \tau }{\mu \nu }}^{\rho \tau }+8R_{\phantom{\mu \nu}{\sigma
\rho}}^{\mu \nu }R_{\phantom {\sigma \kappa} {\nu \tau}}^{\sigma \kappa }R_{%
\phantom{\rho \tau}{ \mu \kappa}}^{\rho \tau }+24R^{\mu \nu \sigma \kappa
}R_{\sigma \kappa \nu \rho }R_{\phantom{\rho}{\mu}}^{\rho }  \nonumber \\
&&+3RR^{\mu \nu \sigma \kappa }R_{\sigma \kappa \mu \nu
}-12RR_{\mu \nu }R^{\mu \nu }+24R^{\mu \nu \sigma \kappa
}R_{\sigma \mu }R_{\kappa \nu }+16R^{\mu \nu }R_{\nu \sigma
}R_{\phantom{\sigma}{\mu}}^{\sigma} +R^{3}. \label{L3}
\end{eqnarray}%
In addition, $\alpha _{i}$'s are Lovelock coefficients and $\mathcal{L}_{%
\mathcal{F}}=\frac{d\mathcal{L}(\mathcal{F})}{d\mathcal{F}}$. Now, we
consider the following line element to obtain the $(n+1)$-dimensional static
topological black hole solutions:
\begin{equation}
ds^{2}=-f(r)dt^{2}+\frac{dr^{2}}{f(r)}+r^{2}d\breve{g}^{2}  \label{Metric}
\end{equation}%
where $d\breve{g}^{2}$ is the metric of an $(n-1)$-dimensional hypersurface
with constant curvature $(n-1)(n-2)k$ and volume $V_{n-1}$ with the
following explicit form
\begin{equation}
d\breve{g}^{2}=\left\{
\begin{array}{cc}
d\theta _{1}^{2}+\sum\limits_{i=2}^{n-1}\prod\limits_{j=1}^{i-1}\sin
^{2}\theta _{j}d\theta _{i}^{2} & k=1 \\
d\theta _{1}^{2}+\sinh ^{2}\theta _{1}d\theta _{2}^{2}+\sinh ^{2}\theta
_{1}\sum\limits_{i=3}^{n-1}\prod\limits_{j=2}^{i-1}\sin ^{2}\theta
_{j}d\theta _{i}^{2} & k=-1 \\
\sum\limits_{i=1}^{n-1}d\phi _{i}^{2} & k=0%
\end{array}%
\right. .  \label{met2}
\end{equation}%
Since the boundary of these spacetimes may be positive, zero or negative
constant curvature, these metric is usually called topological spacetime. At
first we consider the electromagnetic equation (\ref{BIeq}), to obtain the
nonzero component of the gauge potential
\begin{equation}
A_{\mu }=\delta _{\mu }^{0}\times \left\{
\begin{array}{cc}
-\frac{{\beta r\sqrt{{L}_{W}}}}{{2(n-2)(3n-4)}}\left( {(n-1)\,\zeta {L}%
_{W}+3n-4}\right) , & \text{ENED} \\
\frac{{2{\beta ^{2}}{r^{n}}}}{{nq}}\left( \eta {-1}\right) , & \text{LNED}%
\end{array}%
\right. ,  \label{Amu1}
\end{equation}%
where $q$ is an integration constant which is related to the electric charge
and
\begin{equation}
{L}_{W}=LambertW\left( \frac{4q^{2}}{\beta ^{2}r^{2d-4}}{)}\right)
\label{Lw}
\end{equation}%
\begin{equation}
\zeta ={}_{2}{F_{1}}\left( {[1],\,\left[ {\frac{{5n-6}}{{2(n-1)}}}\right] ,\,%
}\frac{{L}_{W}}{{2(n-1)}}\right) ,\,  \label{Fzeta}
\end{equation}

\begin{equation}
\eta ={}_{2}{F_{1}}\left( {\left[ {-\frac{1}{2},\,\frac{{-n}}{{2(n-1)}}}%
\right] ,\,\left[ {\frac{{n-2}}{{2(n-1)}}}\right] ,\,1-}\Gamma ^{2}\right) .
\label{Feta}
\end{equation}%
\begin{equation}
\Gamma =\sqrt{{1}+\frac{{q^{2}}}{{\beta ^{2}}{r^{2(n-1)}}}}
\end{equation}%
It was shown that the mentioned gauge potential reduce to that of Maxwell
field for weak field limit $\beta \rightarrow \infty $. Considering a
special case ${\alpha _{3}=}\frac{{{\alpha ^{2}}}}{{3(n-2)(n-3)(n-4)(n-5)}}$
and ${\alpha _{2}}=\frac{\alpha }{{(n-2)(n-3)}}$, we can show that the
metric function%
\begin{equation}
f(r)=k+\frac{r^{2}}{\alpha }\left( 1-H^{1/3}\right) {,}  \label{fr}
\end{equation}%
with%
\begin{equation}
H=1+\frac{{3\alpha m}}{{{r^{n}}}}+\frac{{6\alpha \Lambda }}{{n(n-1)}}+\frac{{%
3\alpha {\beta ^{2}}}}{{n(n-1){r^{n}}}}\times \left\{
\begin{array}{cc}
{{r^{n}}}+\frac{{2nq}}{{\beta }}{\int }\left( {{\sqrt{Lw}-{\frac{{1}}{\sqrt{%
Lw}}}}}\right) {dr}, & \text{ENED} \\
-{8{r^{n}}}+8{n}\int {\,{r^{n-1}}}\left[ \Gamma -{\ln \left( \frac{\Gamma {+1%
}}{2}\right) }\right] {dr}, & \text{LNED}%
\end{array}%
\right. ,  \label{gr}
\end{equation}%
satisfies all components of the field equations (\ref{Geq}). The parameter $%
m $ is an integration constant which is related to finite mass as \cite{ADM}
\begin{equation}
M=\frac{V_{n-1}(n-1)m}{16\pi }.  \label{Mass}
\end{equation}%
We should note that since computing the total mass leads to an infinite
quantity, one may solve this problem by using of background subtraction
method whose asymptotical geometry matches that of the solutions. Another
approach comes from the fact that adding an additional surface action does
not alter the bulk equations of motion. It is known as AdS/CFT inspired
counterterm method \cite{Counterterm}. All methods have the same result and
one may obtain the finite mass (\ref{Mass}) (see appendix for more details).

Now, we should discuss the existence of singularity(ies). To do so, it is
usual to calculate the Kretschmann scalar. It is easy to find that the
Kretschmann is
\begin{equation}
R_{\alpha \beta \gamma \delta }R^{\alpha \beta \gamma \delta }={{f^{\prime
\prime 2}(r)}}+2(n-1){\left( {\frac{{f^{\prime }(r)}}{{r}}}\right) ^{2}}%
+2(n-1)(n-2){\left( {\frac{{f(r)-k}}{{{r^{2}}}}}\right) ^{2}}.  \label{RR}
\end{equation}%
Taking into account the metric function with Eq. (\ref{RR}), one finds the
Kretschmann diverges at $r=0$ and is finite for $r\neq 0$. In order to
interpret the curvature singularity as a black hole, we should look for the
horizon. The horizon(s) is (are) located at the root(s) of $g^{rr}=f(r)=0$.
Numerical calculations shows that, depending on the values of $\alpha $ and $%
\beta $, the metric function has two real positive roots, one extreme root,
one non-extreme root or it may be positive definite (for more details see
\cite{HendiJHEP,HendiAnn}). Hence obtained solutions may be interpreted as
the black holes with two horizons, extreme black holes, Schwarzschild-like
black holes (one non-extreme horizon) or naked singularity. Moreover, using
the series expansion of metric function for large $r$, one finds%
\begin{eqnarray}
\left. f(r)\right\vert _{\text{large\ }r} &=&k-\frac{2\Lambda \left[
n(n-1)-2\alpha \Lambda \right] r^{2}}{n^{2}(n-1)^{2}}-\frac{m\left[
n(n-1)-4\alpha \Lambda \right] }{n(n-1)r^{n-2}}+\frac{2q^{2}\left[
n(n-1)-4\alpha \Lambda \right] }{n(n-1)^{2}(n-2)r^{2n-4}}+\frac{\alpha m^{2}%
}{r^{2n-2}}  \label{flarge} \\
&&-\frac{4\alpha mq^{2}}{(n-1)(n-2)r^{3n-4}}+\frac{4\alpha q^{4}}{%
(n-1)^{2}(n-2)^{2}r^{4n-6}}-\frac{4\left[ n(n-1)-4\alpha \Lambda \right]
q^{4}}{\Upsilon n(3n-4)\beta ^{2}r^{4n-6}}+O(\frac{1}{r^{5n-6}}),  \nonumber
\\
&&  \nonumber \\
\Upsilon &=&\left\{
\begin{array}{cc}
2(n-1)^{2} & ENED \\
n^{2} & LNED%
\end{array}%
\right. .  \nonumber
\end{eqnarray}%
Eq. (\ref{flarge}) shows that the second term is dominant for large $r$ in
which confirms that these black holes are asymptotically adS if we replace $%
\Lambda $ with $\Lambda _{eff}=\frac{\Lambda \left[ n(n-1)-2\alpha \Lambda %
\right] }{n(n-1)}$. In other words, Lovelock gravity may modify the
cosmological constant and, as we expect, $\Lambda _{eff}\longrightarrow
\Lambda $ for vanishing $\alpha $.

\begin{figure}[tbp]
$%
\begin{array}{cc}
\epsfxsize=5.3cm \epsffile{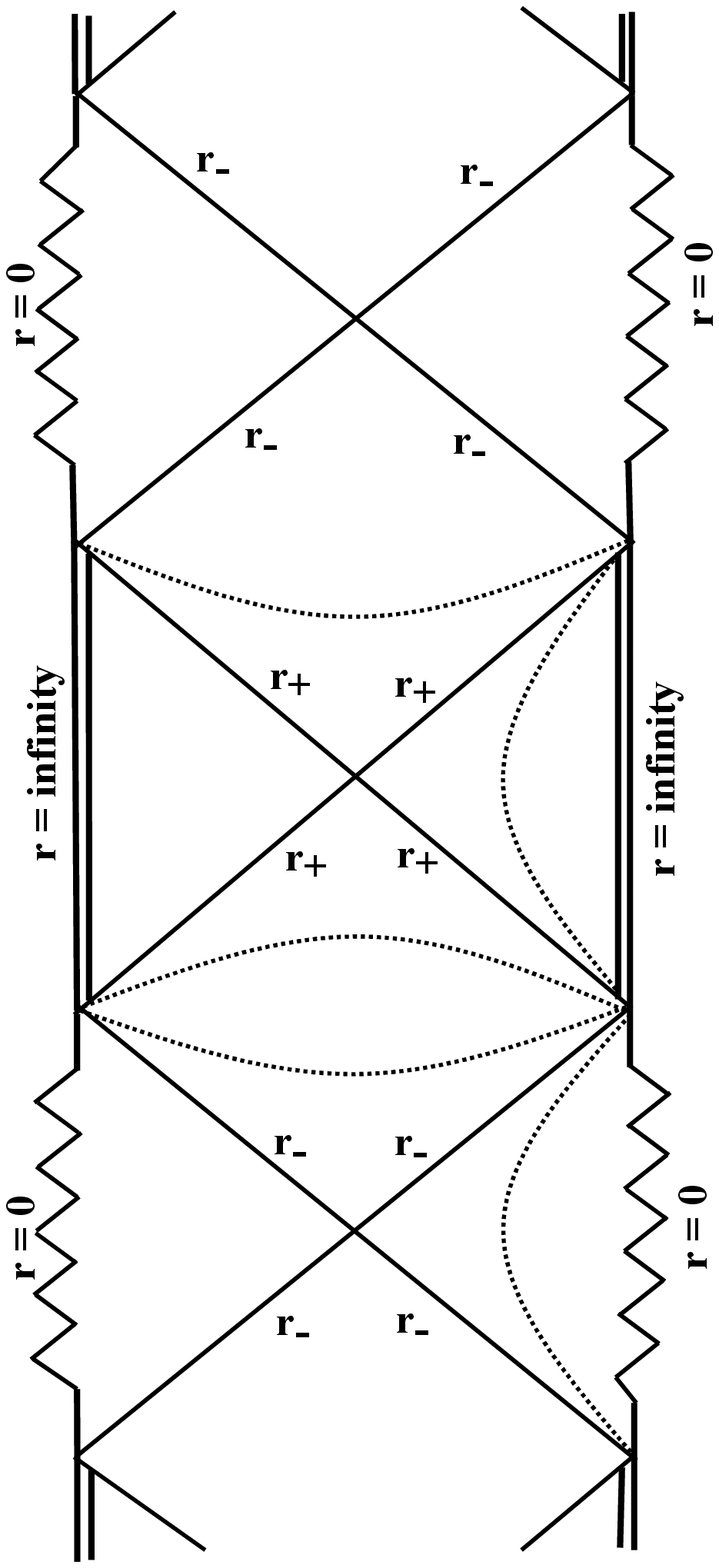} & \epsfxsize=8cm %
\epsffile{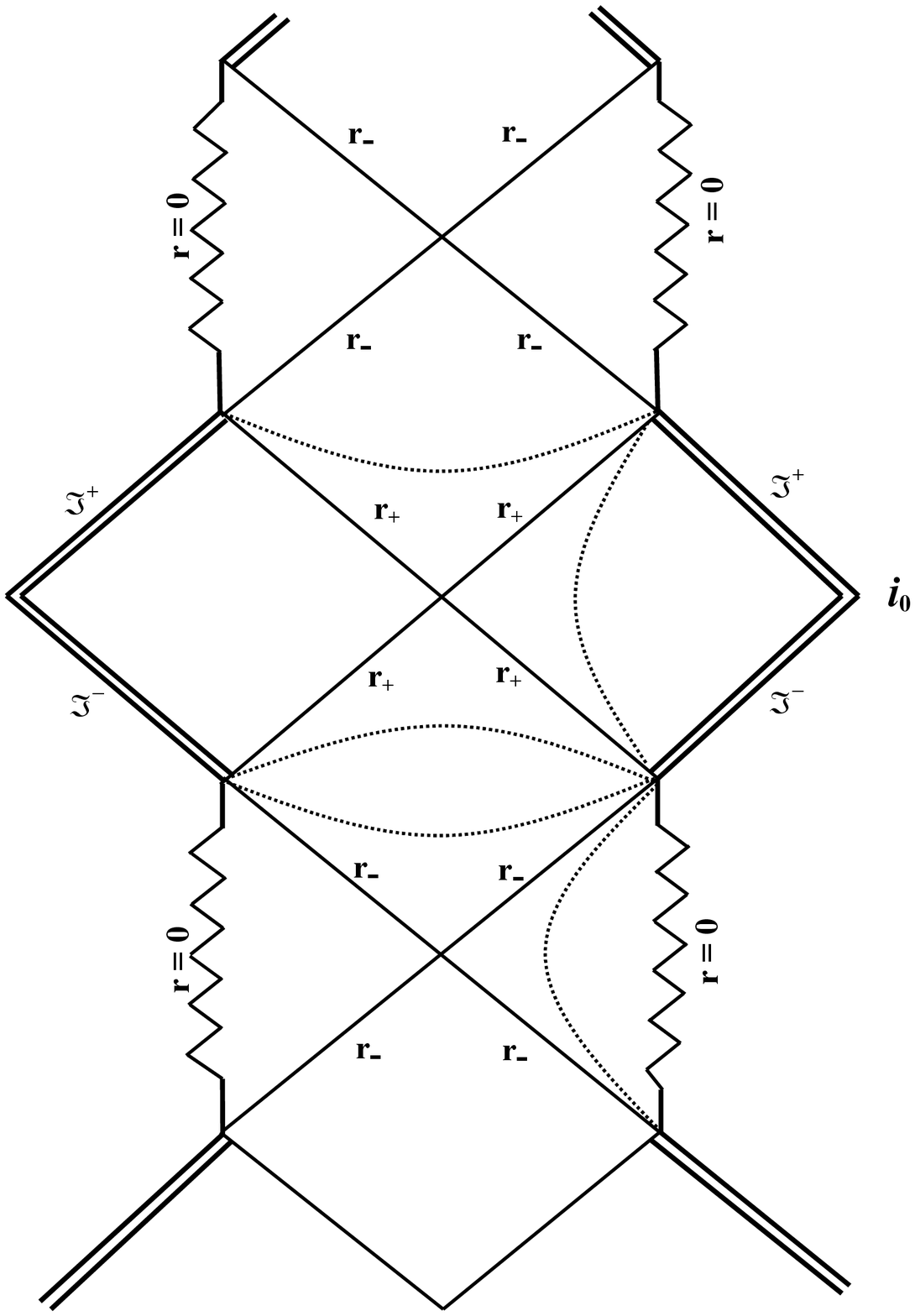}%
\end{array}
$%
\caption{ Carter-Penrose diagram for the asymptotically adS (left figure)
and the asymptotically flat (right figure) black holes when the metric
function has two real positive roots ($r_{-}$ and $r_{+}$). "dotted curves
represent $r=$ constant" }
\label{Pen1}
\end{figure}

\begin{figure}[tbp]
$%
\begin{array}{cc}
\epsfxsize=4.3cm \epsffile{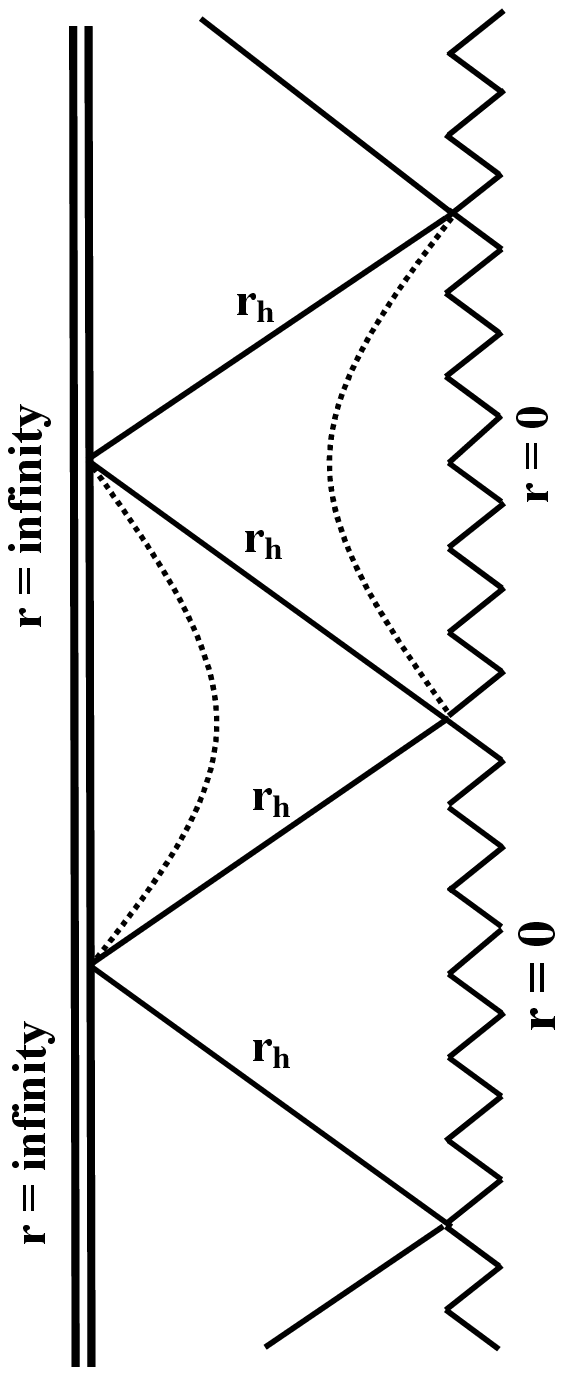} & \epsfxsize=5.5cm %
\epsffile{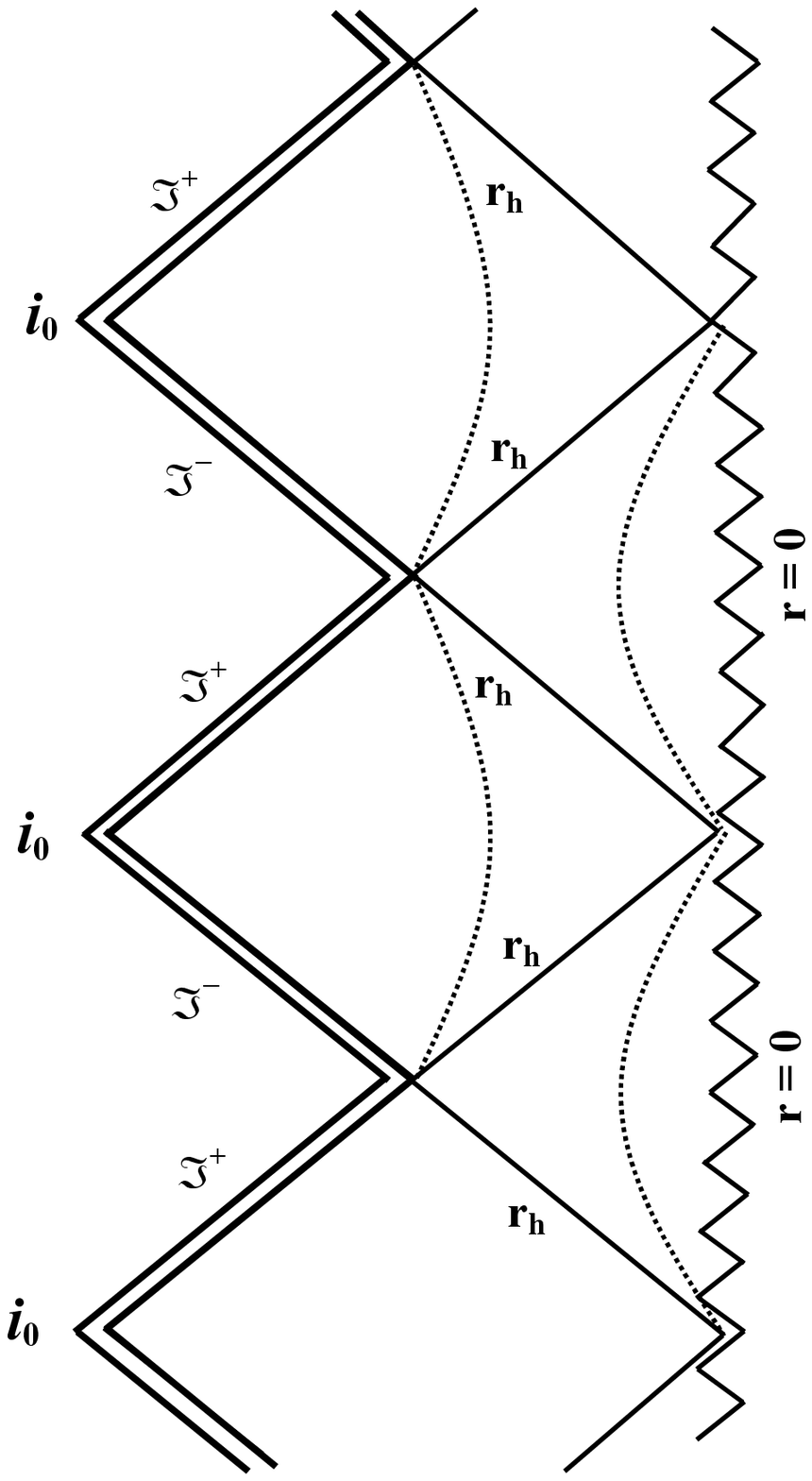}%
\end{array}
$%
\caption{ Carter-Penrose diagram for the asymptotically adS (left figure)
and the asymptotically flat (right figure) black holes when the metric
function has one real positive extreme root ($r_{-}=r_{+}=r_{h}$) (extreme
black hole). "dotted curves represent $r=$ constant" }
\label{Pen2}
\end{figure}

\begin{figure}[tbp]
$%
\begin{array}{cc}
\epsfxsize=5.5cm \epsffile{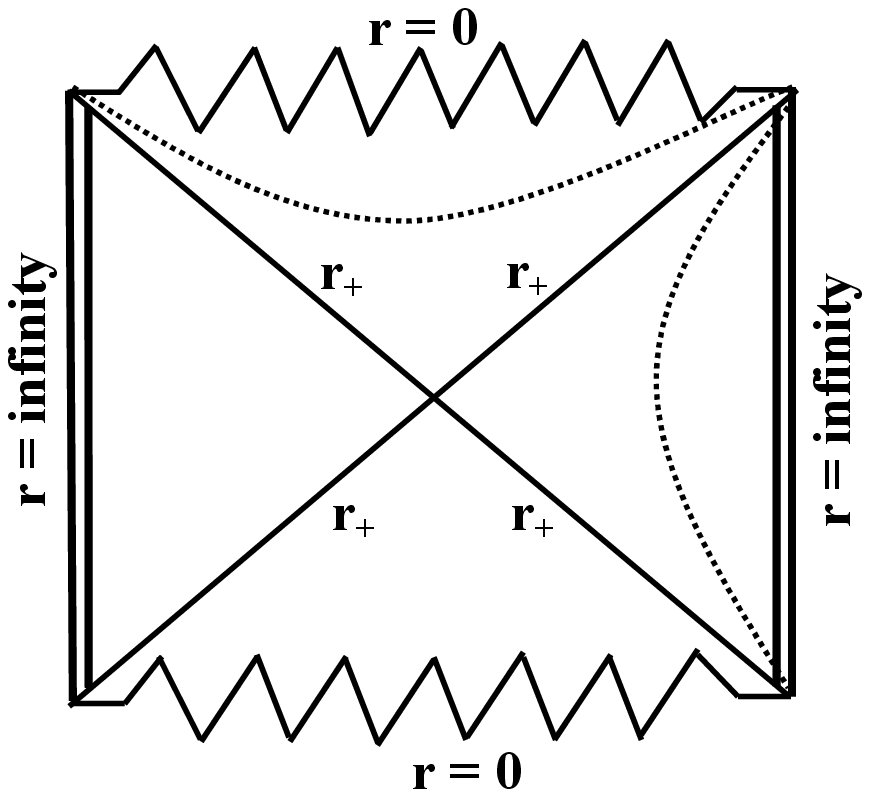} & \epsfxsize=8.5cm %
\epsffile{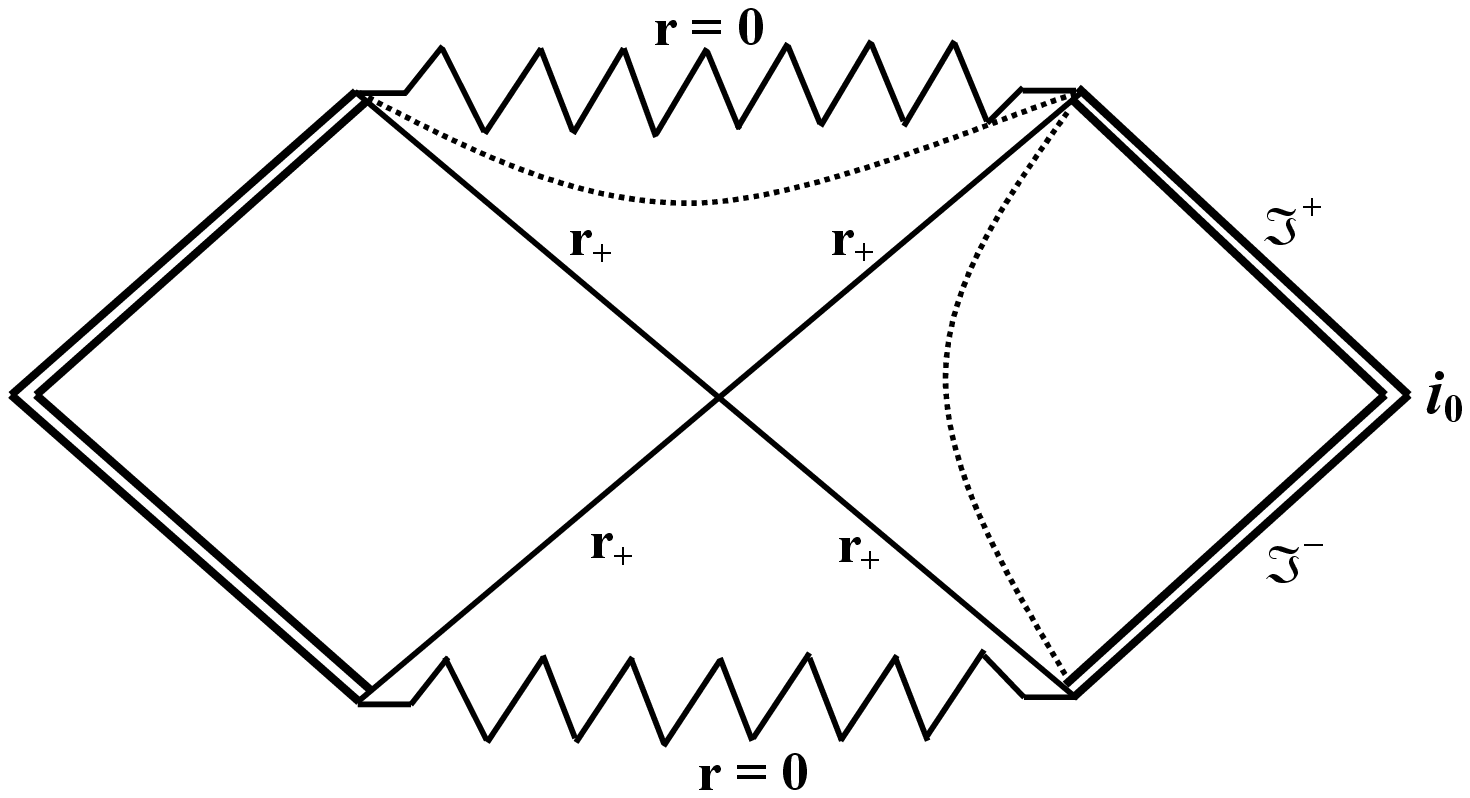}%
\end{array}
$%
\caption{ Carter-Penrose diagram for the asymptotically adS (left figure)
and the asymptotically flat (right figure) black holes when the metric
function has one real positive root ($r_{+}$) (the same as Schwarzschild
black hole). "dotted curves represent $r=$ constant"}
\label{Pen3}
\end{figure}

In order to provide additional information for the conformal
structure of the solutions, we can use the conformal
compactification method to draw the Carter-Penrose (conformal)
diagram (see Figs. \ref{Pen1}-\ref{Pen3} for more details). The
Carter-Penrose diagrams confirm that the singularity may be
timelike (such as Reissner-Nordstr\"{o}m black holes) or spacelike
(such as Schwarzschild black holes). In other words, depending the
values of the nonlinearity parameter, one can find that $\lim_{r
\longrightarrow 0}f(r)$ can be positive or negative (for more
details we refer the reader to Ref. \cite{HendiJHEP,HendiAnn}).
Drawing the Carter-Penrose diagrams shows that the causal
structure of the solutions are asymptotically well behaved.

\section{Thermodynamic properties and thermal stability \label%
{Thermodynamics}}

In this section, we calculate the conserved and thermodynamic quantities,
and check the first law of black hole thermodynamics. Then we perform the
stability criterion.

At first, we apply the definition of surface gravity to obtain the Hawking
temperature
\begin{equation}
T=\frac{1}{2\pi }\sqrt{-\frac{1}{2}\left( \nabla _{\mu }\chi _{\nu }\right)
\left( \nabla ^{\mu }\chi ^{\nu }\right) },  \label{T1}
\end{equation}%
where $\chi $ is the temporal Killing vector, $\partial _{t}$. One obtains%
\begin{equation}
T=\frac{f^{\prime }(r_{+})}{4\pi }=\frac{(n-1)k\left[ 3{\left( {n-2}\right)
r_{+}^{4}+3\left( {n-4}\right) {k}\alpha {r_{+}^{2}}+\left( {n-6}\right) {%
\alpha ^{2}}}\right] -6{\Lambda }r_{+}^{6}+3{{\beta ^{2}}r_{+}^{6}}\Psi }{{%
12\pi (n-1){r_{+}{\left( {r_{+}^{2}+k\alpha }\right) }^{2}}}},  \label{T2}
\end{equation}%
where%
\begin{equation}
\Psi =\left\{
\begin{array}{cc}
\frac{{2\left( {1-}L_{W+}\right) }}{\sqrt{L_{W+}}}\sqrt{\Gamma _{+}^{2}-1}-1,
& \text{ENED} \\
8\left[ 1+{\ln \left( \frac{1+\Gamma _{+}}{2}\right) }-\Gamma _{+}\right] ,
& \text{LNED}%
\end{array}%
\right. ,
\end{equation}%
\begin{equation}
\Gamma _{+}={\sqrt{{1}+\frac{{q^{2}}}{{\beta ^{2}}{r_{+}^{2(n-1)}}}},}
\end{equation}
which shows that the temperature depends on the Lovelock parameter as well
as nonlinearity factor of electrodynamics.

Now, we calculate the entropy of the black hole solutions. Since we regard
the Lovelock gravity, the area law of the Black hole entropy does not
satisfy in general \cite{Entropyfail}. The expression of the entropy of
Lovelock black holes may be derived by Hamiltonian formalism \cite%
{EntropyLov1} (for its generalization for arbitrarily high order derivatives
of the curvature, see \cite{EntropyLov2}), yielding
\begin{equation}
S=\frac{{{V_{n-1}}}}{4}r_{+}^{n-1}\left( {1+\frac{{2\left( {n-1}\right) k}}{{%
\left( {n-3}\right) }}\frac{\alpha }{{r_{+}^{2}}}+\frac{{\left( {n-1}\right)
{k^{2}}}}{{\left( {n-5}\right) }}\frac{{{\alpha ^{2}}}}{{r_{+}^{4}}}}\right)
,  \label{Entropy}
\end{equation}%
It is clear that Eq. (\ref{Entropy}) reproduces the area law for Einstein
gravity ($\alpha \longrightarrow 0$).

In order to obtain the electric charge, we calculate the flux of the
electromagnetic field at infinity. It is easy to show that
\begin{equation}
Q=\frac{V_{n-1}}{4\pi }q.  \label{Charge}
\end{equation}%
Eq. (\ref{Charge}) confirms that the electric charge does not depend on the
nonlinearity parameter.

In order to calculate the electric potential of the black holes, one should
consider a reference. Considering nonzero component of the gauge potential
(or the electric field), one finds that for $r\longrightarrow \infty $, both
the gauge potential and the electric field vanishes. Therefore, it is
natural to calculate the electric potential of the event horizon of black
holes, $r_{+}$, with respect to the infinity as reference \cite{Potential}.
We obtain
\begin{eqnarray}
\Phi &=&A_{\mu }\chi ^{\mu }\left\vert _{r\rightarrow \infty }-A_{\mu }\chi
^{\mu }\right\vert _{r=r_{+}}=  \nonumber \\
&&\left\{
\begin{array}{cc}
\frac{{\beta {r_{+}}\sqrt{L{w_{+}}}}}{{2(n-2)(3n-4)}}\left[ {(n-1)\,{\zeta
_{+}L}_{W+}+3n-4}\right] , & \text{ENED} \\
-\frac{{2{\beta ^{2}}{r_{+}}^{n}}}{{nq}}\left( {{\eta _{+}}-1}\right) , &
\text{LNED}%
\end{array}%
\right. .  \label{Pot}
\end{eqnarray}

\begin{equation}
\zeta _{+}={}_{2}{F_{1}}\left( {[1],\,\left[ {\frac{{5n-6}}{{2(n-1)}}}\right]
,\,}\frac{{L}_{W+}}{{2(n-1)}}\right) ,
\end{equation}

\begin{equation}
\eta _{+}={}_{2}{F_{1}}\left( {\left[ {-\frac{1}{2},\,\frac{{-n}}{{2(n-1)}}}%
\right] ,\,\left[ {\frac{{n-2}}{{2(n-1)}}}\right] ,\,1-}\Gamma
_{+}^{2}\right) ,
\end{equation}

Here, we are in a position to check the first law of thermodynamics for
various horizon topology. At first we obtain the finite mass $M$ as a
function of the entropy and electric charge as the extensive quantities.
Straightforward calculations show that
\begin{equation}
M\left( S,Q\right) =\frac{(n-1){{r_{+}^{n}}}}{{48\pi \alpha }}\left[ \left(
1+\frac{k\alpha }{r_{+}^{2}}\right) ^{3}-1\right] -\frac{{\Lambda {r_{+}^{n}}%
}}{8\pi {n}}-\Theta ,  \label{Smark1}
\end{equation}%
where%
\begin{equation}
\Theta =\left\{
\begin{array}{cc}
\frac{{{\beta ^{2}r_{+}^{n}}}}{16\pi {n}}+\frac{{\beta q}}{8\pi }\left. {%
\int }\left( {{\sqrt{Lw}-{\frac{{1}}{\sqrt{Lw}}}}}\right) {dr}\right\vert
_{r_{+}}, & \text{ENED} \\
-\frac{{{\beta ^{2}r_{+}^{n}}}}{2\pi {n}}+\frac{{\beta }^{2}}{2\pi }\left.
\int {\,{r^{n-1}}}\left[ \Gamma -{\ln \left( \frac{\Gamma {+1}}{2}\right) }%
\right] {dr}\right\vert _{r_{+}}, & \text{LNED}%
\end{array}%
\right. .
\end{equation}

Now, we use the first law to define temperature and electric potential as
the intensive parameters conjugate to the entropy and electric charge
\begin{eqnarray}
&&T=\left( \frac{\partial M}{\partial S}\right) _{Q}=\frac{\left( \frac{%
\partial M}{\partial r_{+}}\right) _{Q}}{\left( \frac{\partial S}{\partial
r_{+}}\right) _{Q}},  \label{Tk} \\
&&\Phi =\left( \frac{\partial M}{\partial Q}\right) _{S}=\frac{\left( \frac{%
\partial M}{\partial q}\right) _{r_{+}}}{\left( \frac{\partial Q}{\partial q}%
\right) _{r_{+}}}.  \label{Phik}
\end{eqnarray}

Numerical analysis shows that Eqs. (\ref{Tk}) and (\ref{Phik}) are equal to
Eqs. (\ref{T2}) and (\ref{Pot}), respectively, and therefore we deduce that
these quantities satisfy the first law of thermodynamics
\begin{equation}
dM=TdS+\Phi dQ.  \label{FirstLaw}
\end{equation}

\begin{figure}[tbp]
$%
\begin{array}{cc}
\epsfxsize=8cm \epsffile{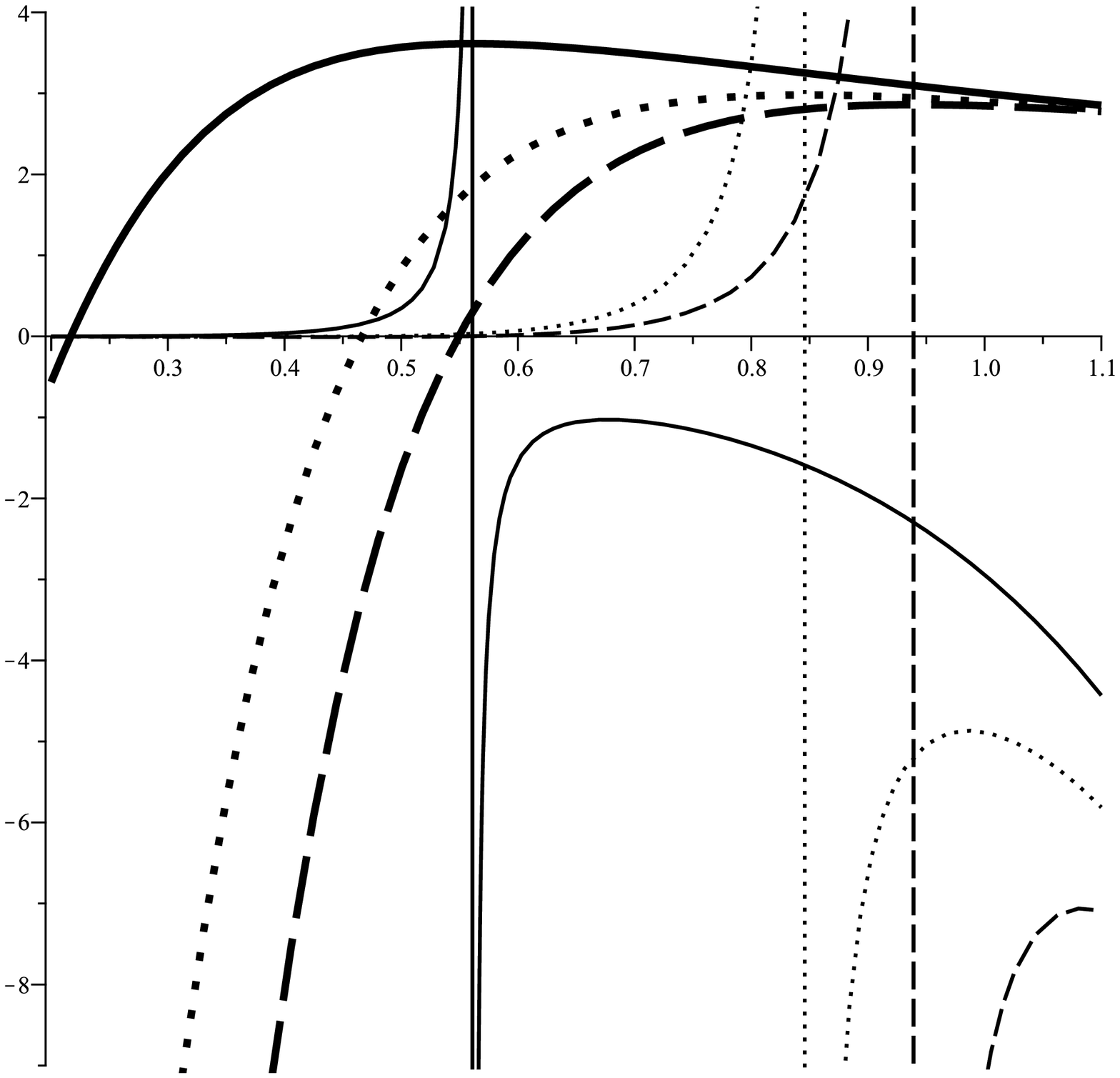} & \epsfxsize=8cm %
\epsffile{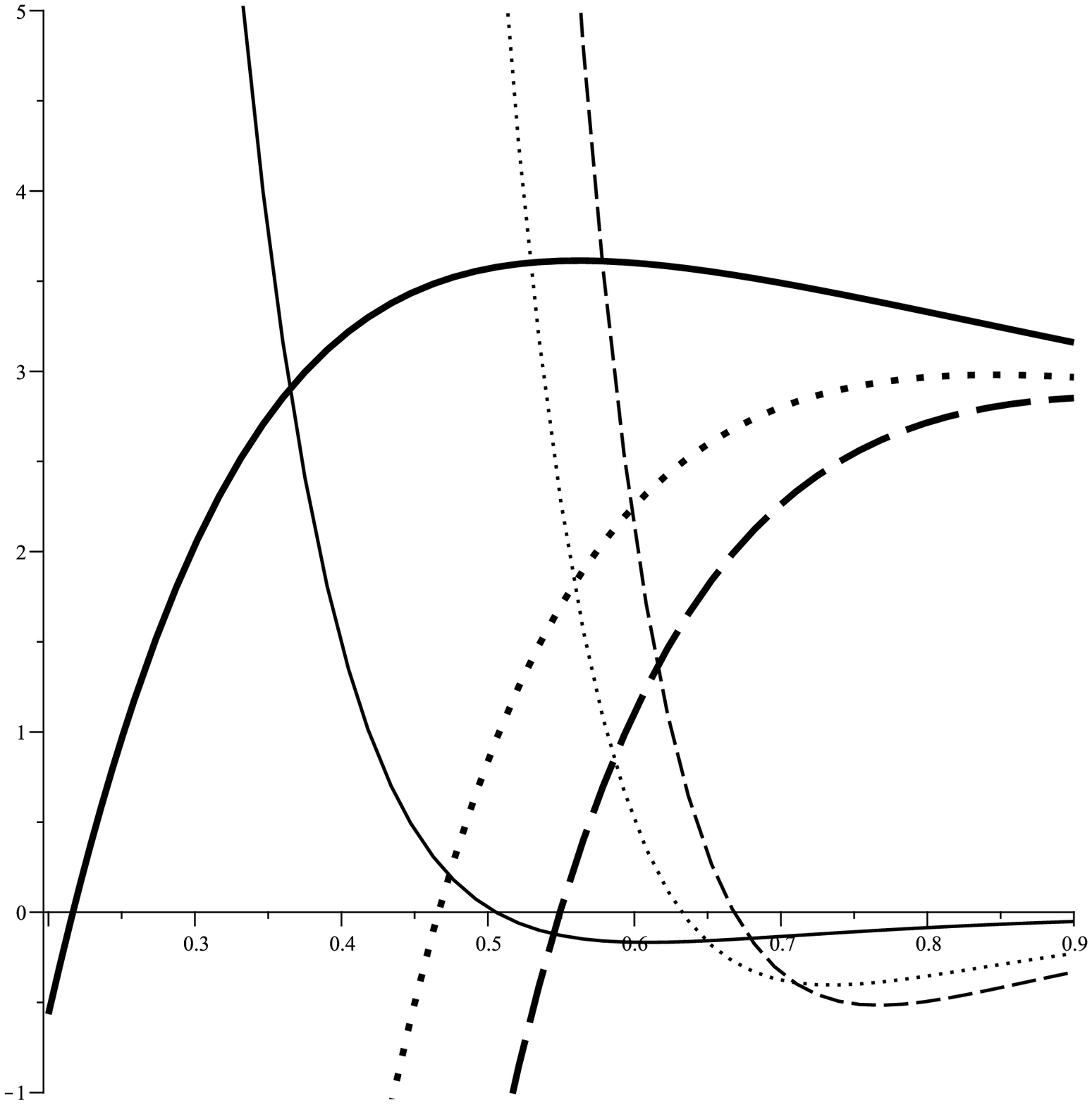}%
\end{array}
$%
\caption{$C_{Q}$ (left), $H$ (right) diagrams versus $r_{+}$ for $k=1$, $n=6$%
, $q=1$, $\protect\alpha=0.1$, and $\protect\beta=0.05$ (solid line) $%
\protect\beta=0.5$ (dotted line) $\protect\beta=1$ (dashed line) \textbf{%
"bold lines are related to corresponding temperatures"} }
\label{CH1}
\end{figure}
\begin{figure}[tbp]
$%
\begin{array}{cc}
\epsfxsize=8cm \epsffile{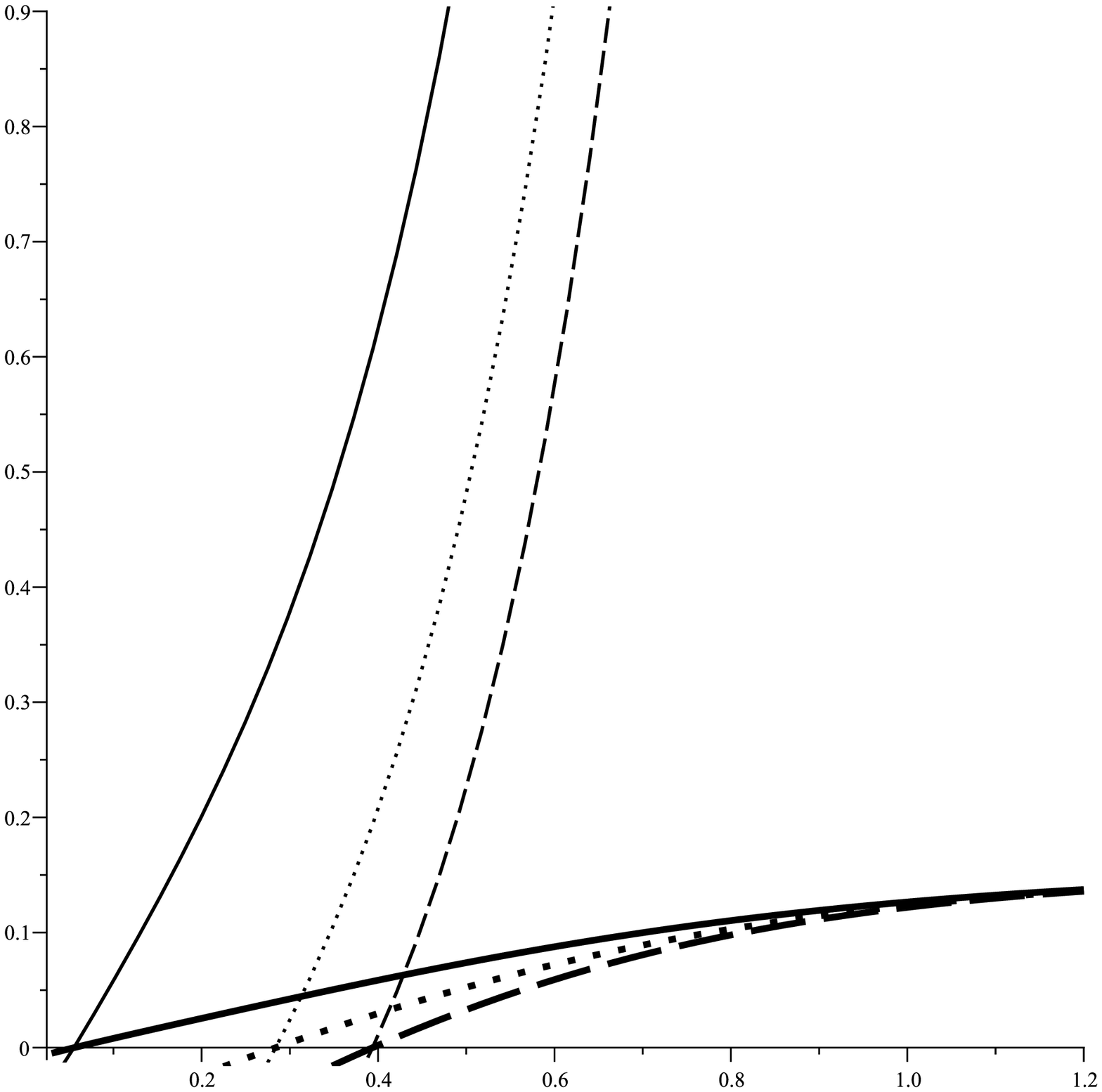} & \epsfxsize=8cm %
\epsffile{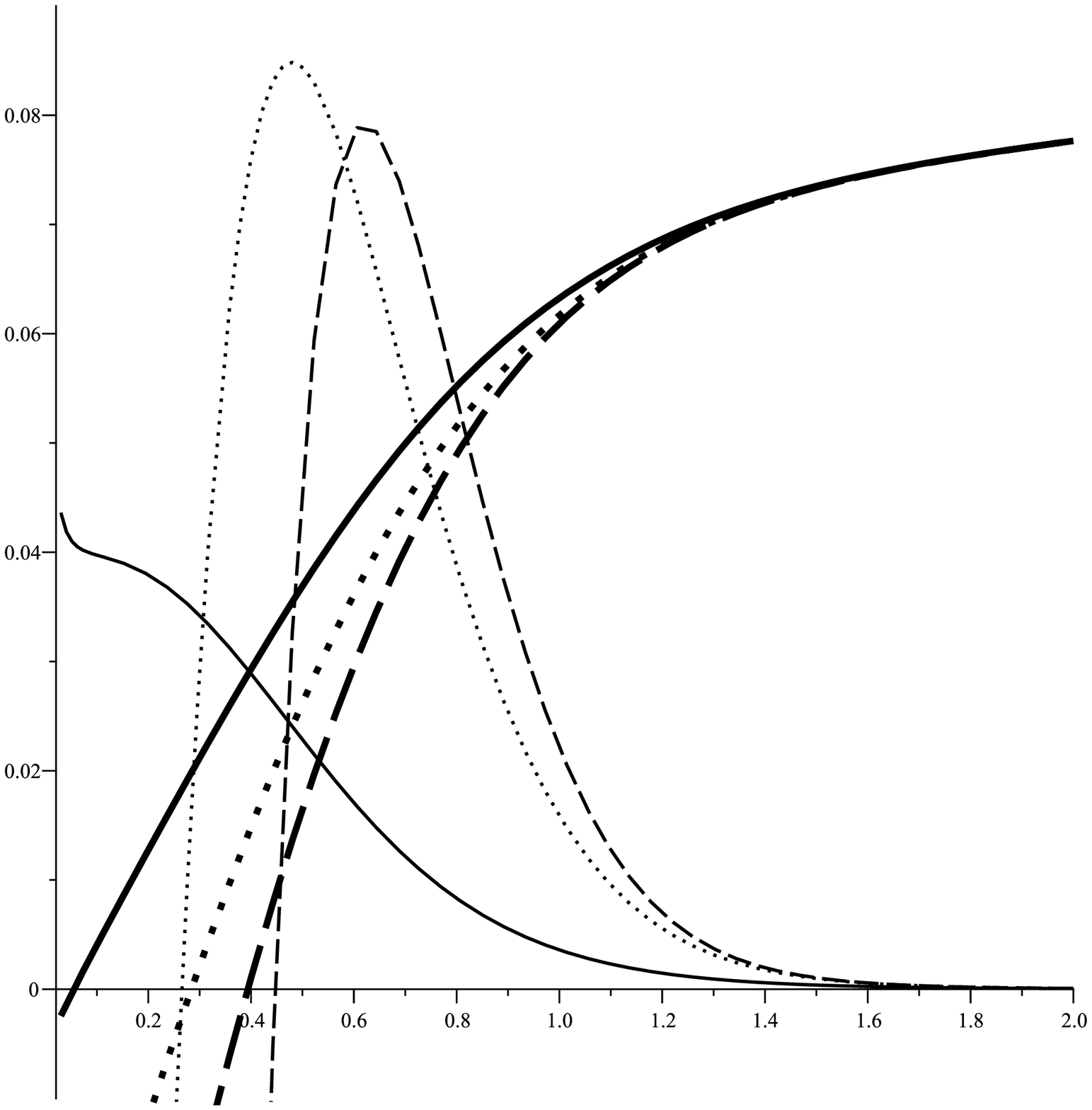}%
\end{array}
$%
\caption{$C_{Q}$ (left), $H$ (right) diagrams versus $r_{+}$ for $k=1$, $n=6$%
, $q=1$, $\protect\alpha=1$, and $\protect\beta=0.05$ (solid line) $\protect%
\beta=0.5$ (dotted line) $\protect\beta=1$ (dashed line) \textbf{"bold lines
are related to corresponding temperatures"} }
\label{CH2}
\end{figure}

\begin{figure}[tbp]
$%
\begin{array}{cc}
\epsfxsize=8cm \epsffile{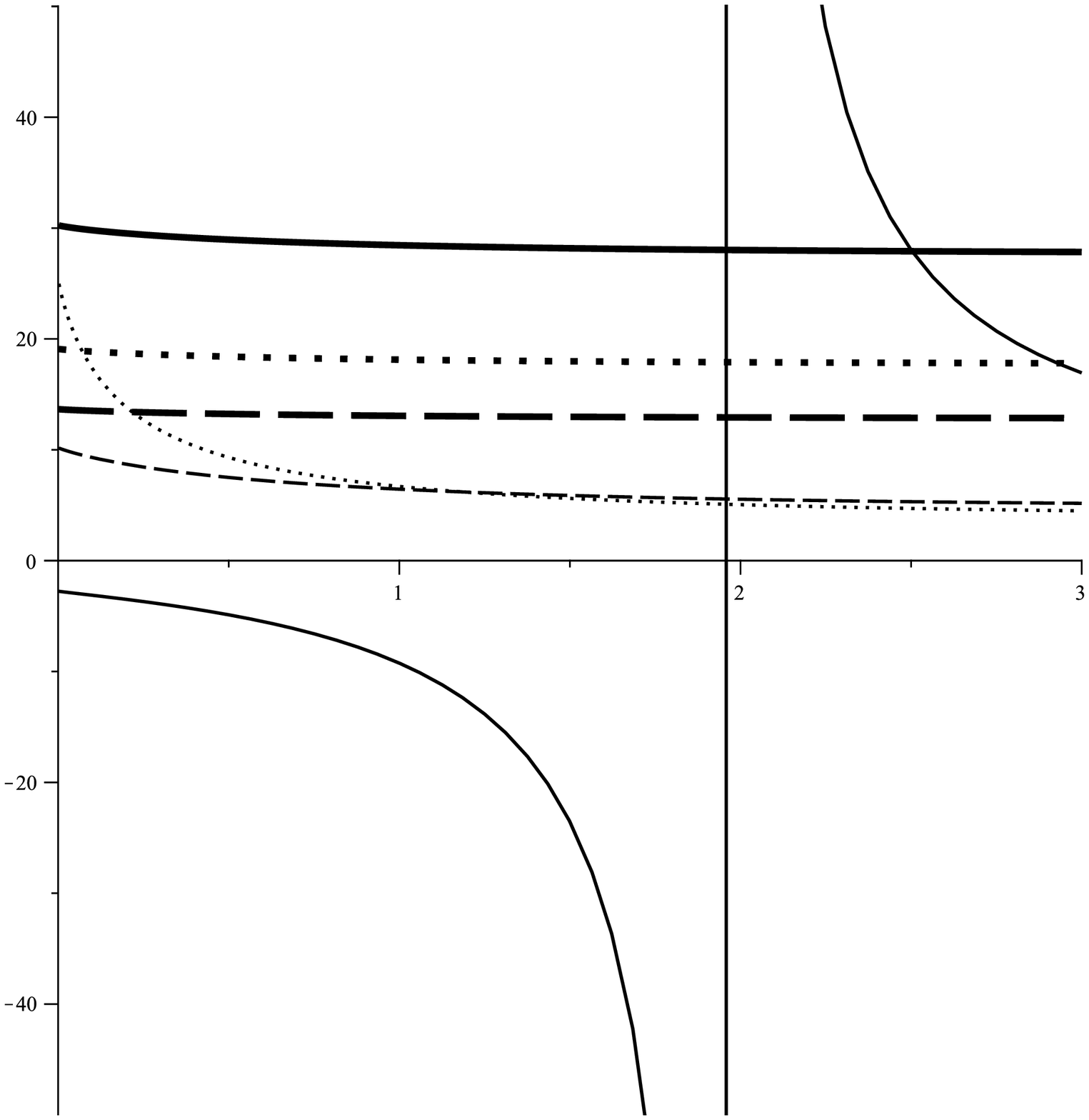} & \epsfxsize=8cm %
\epsffile{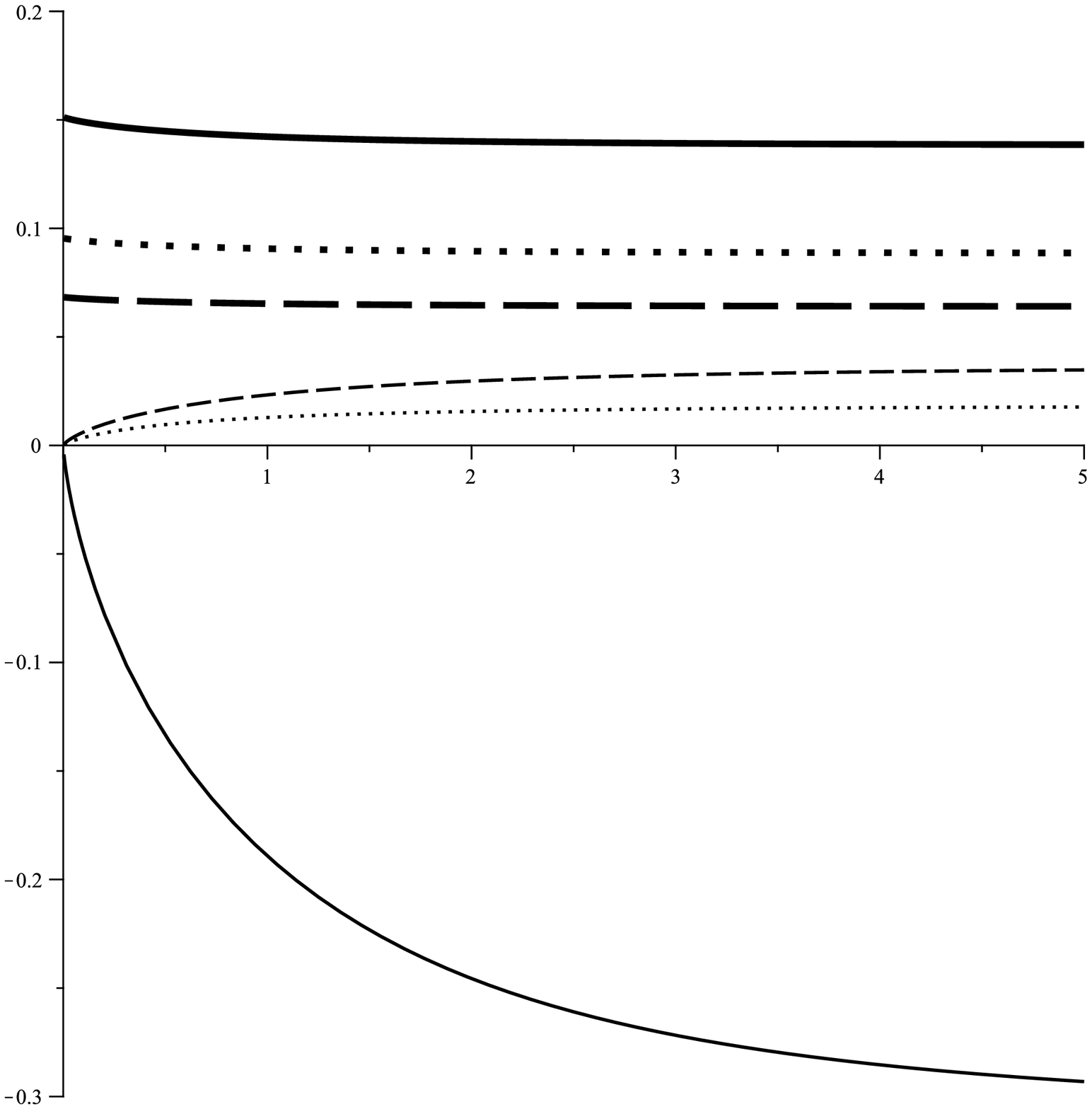}%
\end{array}
$%
\caption{$C_{Q}$ (left), $H$ (right) diagrams versus $\protect\beta$ for $%
k=1 $, $n=6$, $q=1$, $r_{+}=1$, and $\protect\alpha=0.1$ (solid line) $%
\protect\alpha=0.5$ (dotted line) $\protect\alpha=0.9$ (dashed line) \textbf{%
"bold lines are related to corresponding temperatures"} }
\label{CH3}
\end{figure}
\begin{figure}[tbp]
$%
\begin{array}{cc}
\epsfxsize=8cm \epsffile{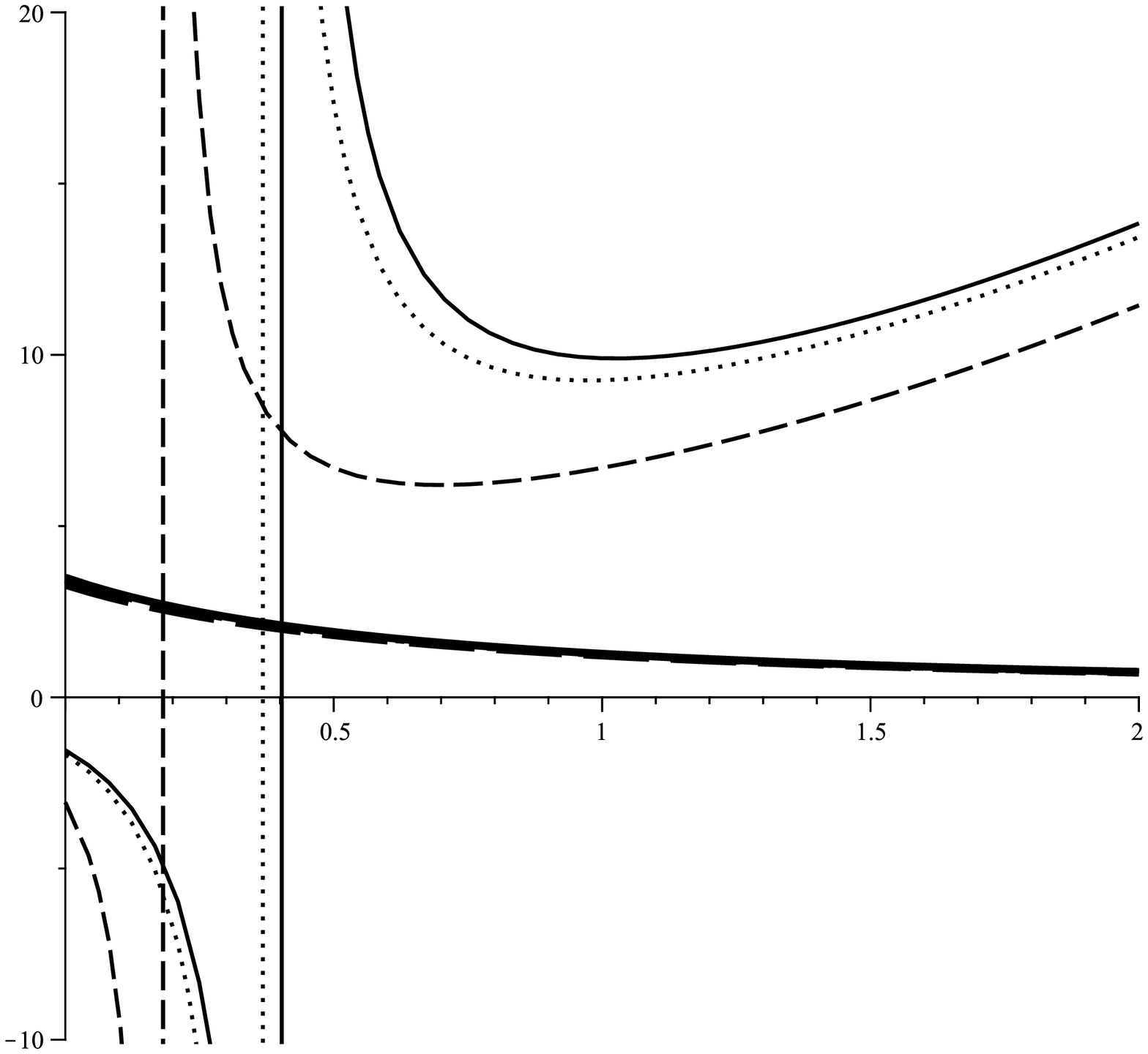} & \epsfxsize=8cm %
\epsffile{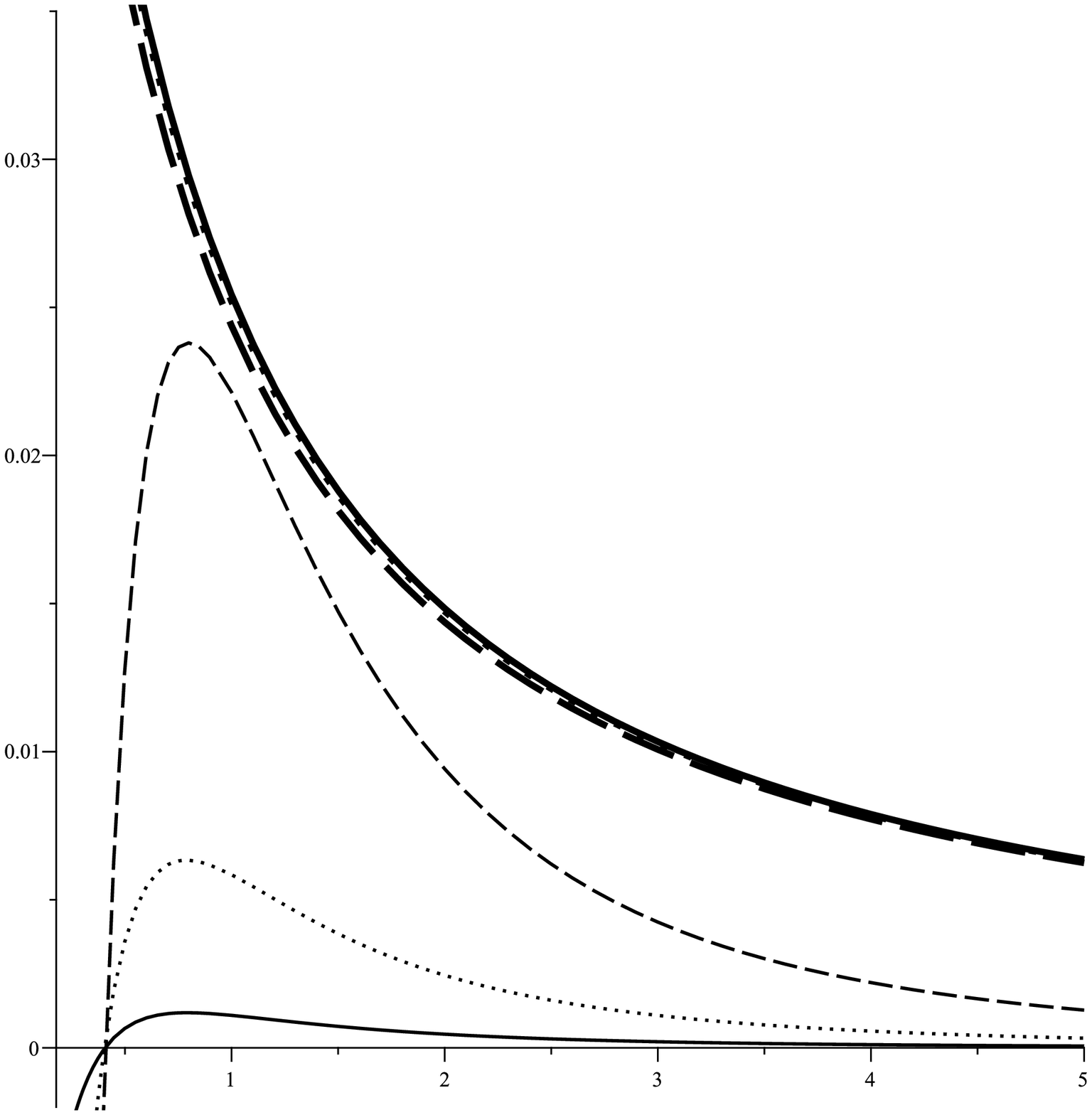}%
\end{array}
$%
\caption{$C_{Q}$ (left), $H$ (right) diagrams versus $\protect\alpha$ for $%
k=1$, $n=6$, $q=1$, $r_{+}=1$, and $\protect\beta=0.05$ (solid line) $%
\protect\beta=0.5$ (dotted line) $\protect\beta=1$ (dashed line) \textbf{%
"bold lines are related to corresponding temperatures"} }
\label{CH4}
\end{figure}

\subsection{Thermal stability\label{Stab}}

In order to discuss the thermal stability conditions, one may use both
canonical and grand canonical ensembles. The positivity of the heat capacity
and determinant of the Hessian matrix are the requirements usually referred
to stability criterion in canonical and grand canonical ensembles,
respectively. Since $M$ is a function of $S$ and $Q$, we can write the heat
capacity and determinant of the Hessian matrix with the following explicit
forms
\begin{eqnarray}
C_{Q} &=&\frac{T}{\left( \frac{\partial ^{2}M}{\partial S^{2}}\right) _{Q}},
\label{CQ} \\
H &=&det\left[
\begin{array}{cc}
\frac{\partial ^{2}M}{\partial S^{2}} & \frac{\partial ^{2}M}{\partial
S\partial Q} \\
\frac{\partial ^{2}M}{\partial Q\partial S} & \frac{\partial ^{2}M}{\partial
Q^{2}}%
\end{array}%
\right] .  \label{H}
\end{eqnarray}

Analytical calculations of the heat capacity and determinant of the Hessian
matrix are too large and therefore we leave out the analytical result for
reasons of economy. We plot some figures to discuss the stability
conditions. Numerical calculations show that although large black holes have
positive definite temperature, there is a lower limit for the horizon radius
of physical small black holes, $r_{0}$. It is notable that $r_{0}$ increases
for increasing $\beta $ (decreasing $\alpha $). In addition, we find that,
for small values of $\alpha $ ($\alpha <\alpha _{c}$), there are two $r_{a}$
and $r_{b}$ ($r_{a}<r_{b}$), in which the black holes are stable for $%
r_{0}<r_{+}<r_{a}$ and $r_{+}>r_{b}$ (see Fig. \ref{CH1} and also following
three tables). Moreover, for large values of $\alpha $ ($\alpha >\alpha _{c}$%
) and $r_{+}>r_{c}$ the black holes are stable. We should note that for
canonical ensemble one finds $r_{c}=r_{0}$ and for grand canonical ensemble $%
r_{c}>r_{0}$ (see Fig. \ref{CH2}). Figs. \ref{CH3} and \ref{CH4} confirm
that, regardless of value of $\beta $, we encounter an unstable phase for $%
\alpha <\alpha _{c}$.

\vspace{0.1cm}

\begin{center}
\begin{tabular}{rccccc}
\hline\hline
$r_{+}=$ & \vline & $0.5$ & $1$ & $4$ & $400$ \\ \hline
$C_{Q}=$ & \vline & $0.35$ & $-2.9$ & $5148$ & $1.2\times 10^{13}$ \\ \hline
$H=$ & \vline & $0.02$ & \hspace{0.3cm} $-0.03$ & \hspace{0.5cm} $4.8\times
10^{-5}$ & \hspace{0.5cm} $1.2\times 10^{-20}$ \\ \hline
\end{tabular}
\\[0pt]
\vspace{0.1cm} Table (1): corresponding to Fig. \ref{CH1} for $\beta=0.05$.

\begin{tabular}{rccccc}
\hline\hline
$r_{+}=$ & \vline & $0.6$ & $1$ & $4$ & $400$ \\ \hline
$C_{Q}=$ & \vline & $0.07$ & $-4.8$ & $5148$ & $1.2\times 10^{13}$ \\ \hline
$H=$ & \vline & $0.51$ & \hspace{0.3cm}$-0.13$ & \hspace{0.5cm} $4.8\times
10^{-5}$ & \hspace{0.5cm} $1.2\times 10^{-20}$ \\ \hline
\end{tabular}
\\[0pt]
\vspace{0.1cm} Table (2): corresponding to Fig. \ref{CH1} for $\beta=0.5$.

\begin{tabular}{rccccc}
\hline\hline
$r_{+}=$ & \vline & $0.6$ & $1$ & $4$ & $400$ \\ \hline
$C_{Q}=$ & \vline & $0.02$ & $-9.25$ & $5148$ & $1.2\times 10^{13}$ \\ \hline
$H=$ & \vline & $2.1$ & \hspace{0.3cm} $-0.19$ & \hspace{0.5cm} $4.8\times
10^{-5}$ & \hspace{0.5cm} $1.2\times 10^{-20}$ \\ \hline
\end{tabular}
\\[0pt]
\vspace{0.1cm} Table (3): corresponding to Fig. \ref{CH1} for $\beta=1$.
\end{center}


\section{Extended phase space and Smarr formula}

In previous section we considered the usual discussions of
thermodynamic properties of asymptotically adS black holes, in
which the cosmological constant is treated as a fixed parameter.
However, there are some motivations to view the cosmological
constant as a variable (for e.g. see \cite{LambdaP}). In addition,
there exist some theories where some physical constants such as
gauge coupling constants, Newton constant, Lovelock coefficients
and BI parameter may not be fixed values but dynamical ones. In
that case, it is natural to consider these variable parameters
into the first law of black hole thermodynamics
\cite{CteVariable}. Considering
the cosmological constant as a thermodynamic pressure, the black hole mass $%
M $ should be explained as enthalpy rather than internal energy of the
system \cite{Enthalpy}. In the geometric units, one can identify the
cosmological constant with the pressure as
\begin{equation}
P=-\frac{\Lambda }{8\pi },  \label{P}
\end{equation}%
where the thermodynamic quantity conjugate to the pressure is called
thermodynamic volume of black holes. In addition, it was shown that the
Smarr formula may be extended to Lovelock gravity as well as nonlinear
theories of electrodynamics \cite{SmarrNew}.

Geometrical techniques (scaling argument) were used to derive an
extension of the first law and its related modified Smarr formula
that includes variations in the cosmological constant, Lovelock
coefficient and also nonlinearity parameter. In our case, Lovelock
gravity in the presence of the NED, $M$ should be the function of
entropy, pressure,
charge, Lovelock parameter and BI coupling coefficient \cite%
{SmarrNew}. Regarding the previous section, we find that those thermodynamic
quantities satisfy the following differential form
\begin{equation}
dM=TdS+\Phi dQ+VdP+\mathcal{A}_{1}^{\prime }d\alpha _{2}+\mathcal{A}%
_{2}^{\prime }d\alpha _{3}+\mathcal{B}d\beta .  \label{GenFirstLaw}
\end{equation}%
where we have achieved $T$ and $\Phi $, and one can obtain%
\begin{eqnarray*}
V &=&\left( \frac{\partial M}{\partial P}\right) _{S,Q,\alpha _{2},\alpha
_{3},\beta }, \\
\mathcal{A}_{1}^{\prime } &=&\left( \frac{\partial M}{\partial \alpha _{2}}%
\right) _{S,Q,P,\alpha _{3},\beta }, \\
\mathcal{A}_{2}^{\prime } &=&\left( \frac{\partial M}{\partial \alpha _{3}}%
\right) _{S,Q,P,\alpha _{2},\beta }, \\
\mathcal{B} &=&\left( \frac{\partial M}{\partial \beta }\right)
_{S,Q,P,\alpha _{2},\alpha _{3}}.
\end{eqnarray*}

Using the redefinition of $\alpha _{2}$ and $\alpha _{3}$ with respect to
the single parameter, $\alpha $, we can rewrite $\mathcal{A}_{1}^{\prime
}d\alpha _{2}+\mathcal{A}_{2}^{\prime }d\alpha _{3}$ as a single
differential form
\begin{eqnarray*}
d{\alpha _{2}} &=&\frac{1}{{(n-2)(n-3)}}d\alpha , \\
d\alpha _{3} &=&\frac{2{\alpha }}{{3(n-2)(n-3)(n-4)(n-5)}}{d\alpha .}
\end{eqnarray*}

Moreover, by scaling argument, we can obtain the generalized Smarr relation
for our asymptotically adS solutions in the extended phase space
\begin{equation}
(d-3)M=(d-2)TS+(d-3)Q\Phi -2PV+2\left( \mathcal{A}_{1}\alpha +\mathcal{A}%
_{2}\alpha ^{2}\right) -\mathcal{B}\beta  \label{Smarr2}
\end{equation}%
where
\begin{eqnarray*}
V &=&\frac{r_{+}^{n}}{n}, \\
\mathcal{A}_{1} &=&\frac{(n-1)k^{2}r_{+}^{n-4}}{16\pi }-\frac{%
(n-1)kTr_{+}^{n-3}}{2(n-3)}, \\
\mathcal{A}_{2} &=&\frac{(n-1)k^{3}r_{+}^{n-6}}{24\pi }-\frac{%
(n-1)k^{2}Tr_{+}^{n-5}}{2(n-5)}, \\
\left. \mathcal{B}\right\vert _{\text{ENED}}
&=&\frac{q(n-1)r_{+}\left(
L_{W+}\right) ^{\frac{3}{2}}F\left( [1],[\frac{5n-6}{2n-2}],\frac{L_{W+}}{%
2n-2}\right) }{8\pi n(3n-4)}-\frac{\beta r_{+}^{n}}{8\pi n}+\frac{q\beta
r_{+}^{n+1}\sqrt{L_{W+}}\left( 1-L_{W+}\right) }{8\pi n\left(
1+L_{W+}\right) }+\frac{2qr_{+}}{8\pi n\sqrt{L_{W+}}\left( 1+L_{W+}\right) },
\\
\left. \mathcal{B}\right\vert _{\text{LNED}} &=&\frac{\beta
r_{+}^{n}}{2\pi
n^{2}}\left[ -(n-1)\left( 1-\Gamma _{+}^{2}\right) F\left( \left[ \frac{1}{2}%
,\frac{n-2}{2n-2}\right] ,\left[ \frac{3n-4}{2n-2}\right] ,1-\Gamma
_{+}^{2}\right) +2n\ln \left( \frac{1+\Gamma _{+}}{2}\right) +(3n-2)\left(
1-\Gamma _{+}\right) \right] .
\end{eqnarray*}%
Regarding the mentioned argument and using Eqs. (\ref{P}) and (\ref{T2}),
one can obtain the equation of state $P(V,T)$ to compare the black hole
system with the Van der Waals fluid equation in ($n+1$)-dimensions \cite%
{PVlovelock}.

\section{ CLOSING REMARKS}

In this paper we considered third order Lovelock gravity in the
presence of exponential and logarithmic forms of NED models.
Regardless of naked singularities, we obtained topological black
hole solutions with two horizons or one (non-)extreme horizon. We
found that replacing $\Lambda$ with an effective cosmological
constant, $\Lambda_{eff}$, one may obtain asymptotically adS
solutions. In other words, Lovelock gravity and also BI type NED
models do not alter the asymptotical behavior of the solutions. We
obtained thermodynamics and conserved quantities of the
topological black holes and found that the Lovelock gravity does
not affect the temperature, entropy and finite mass only for black
holes with Ricci flat horizon, $k=0$. Moreover, we showed that the
thermodynamics and conserved quantities satisfy the first law of
thermodynamics.

We performed stability criterion in both canonical and grand canonical
ensembles by use of numerical analysis only for $k=1$. We found a lower
bound for the horizon radius, $r_{0}\geq 0$, in which the temperature is
positive for $r_{+}>r_{0}$. We showed that the nonlinearity parameter, $%
\beta $, and also Lovelock coefficient, $\alpha $ can affect the value of $%
r_{0}$. Then we studied the heat capacity and determinant of Hessian matrix
and showed that for $\alpha <\alpha _{c}$, there are two limits $r_{a}$ and $%
r_{b}$ ($r_{a}<r_{b}$), in which the black holes have an unstable phase for $%
r_{a}<r_{+}<r_{b}$. Furthermore, we found an lower limit ($r_{c}$) in which
for $\alpha >\alpha _{c}$ and $r_{+}>r_{c}$ the black holes are stable. In
addition, we found that for canonical ensemble one finds $r_{c}=r_{0}$ and
for grand canonical ensemble $r_{c}>r_{0}$. Calculations showed that
regardless of value of values of $\beta $, there is an unstable phase of
black hole solutions for $\alpha <\alpha _{c}$.

At last, we have discussed the extended phase space in which the
cosmological constant, nonlinearity and Lovelock parameters
considered as dynamical variables. We have calculated generalized
Smarr formula and also modified first law of thermodynamics.
Extended phase space help us to investigate the similarities
between the thermodynamical behavior of black hole system under
studied and the Van der Waals gas/liquid system.

Finally, we should note that for the sake of economy, we
investigated stability conditions only for $k=1$. One may regard
other horizon topology for discussion of thermal stability. In
addition, it is worthwhile to mention that it would be interesting
to investigate the phase transition by Geometrothermodynamics
approach \cite{GTD}. In addition, one can follow the section IV to
discuss about the concept of extended phase space thermodynamics
and $P-V$ criticality of the Lovelock black holes with BI type NED
\cite{PVlovelock}. We leave these problems to our forthcoming
independent works.

\begin{acknowledgements}
We would like to thank the anonymous referee for useful suggestions and enlightening comments.
The authors wish to thank Shiraz University Research Council. This
work has been supported financially by Center for Excellence in
Astronomy \& Astrophysics of Iran (CEAAI-RIAAM).
\end{acknowledgements}

\begin{center}
\textbf{Appendix}
\end{center}

The action of third order Lovelock gravity in the presence of NED
which is related to the field equations (\ref{Geq}) and
(\ref{BIeq}) is
\begin{equation}
\mathcal{I}_{G}=-\frac{1}{16\pi }\int_{\mathcal{M}}d^{n+1}x\sqrt{-g}%
[R-2\Lambda +\alpha _{2}\mathcal{L}_{2}+\alpha _{3}\mathcal{L}_{3}+\mathcal{L%
}(\mathcal{F})]+\mathcal{I}_{b},  \label{Ibulk}
\end{equation}%
where $\mathcal{L}_{2}$, $\mathcal{L}_{3}$ and $\mathcal{L}(\mathcal{F})$\
were defined before. The last term in Eq. \ref{Ibulk} is boundary action.
The integral of Eq. \ref{Ibulk} does not have a well-defined variational
principle, since one encounters a total derivative that produces a surface
integral involving the derivative of $\delta g_{\mu \nu }$ normal to the
boundary. The normal derivative terms do not vanish by themselves, but are
cancelled by the variation of the suitable surface term
(Gibbons-Hawking-York boundary term \cite{GH1,GH2}) with the following
explicit form%
\begin{equation}
\mathcal{I}_{b}=-\frac{1}{8\pi }\int_{\partial \mathcal{M}}d^{n}x\sqrt{%
-\gamma }\left[ K+\alpha _{2}L_{2b}+\alpha _{3}L_{3b}\right] ,
\label{Iboundary}
\end{equation}%
with%
\begin{equation}
L_{2b}=2\left( J-2\widehat{G}_{ab}^{(1)}K^{ab}\right) ,  \label{L2b}
\end{equation}%
\begin{equation}
L_{3b}=3(P-2\widehat{G}_{ab}^{(2)}K^{ab}-12\widehat{R}_{ab}J^{ab}+2\widehat{R%
}J-4K\widehat{R}_{abcd}K^{ac}K^{bd}-8\widehat{R}%
_{abcd}K^{ac}K_{e}^{b}K^{ed}),  \label{L3b}
\end{equation}%
where $\gamma _{\mu \nu }$ and $K$ are, respectively, the induced metric and
the trace of extrinsic curvature of boundary, $\widehat{G}_{ab}^{(1)}$ and $%
\widehat{G}_{ab}^{(2)}$ denote the $n$-dimensional Einstein and second order
Lovelock tensors of the metric $\gamma _{ab}$ while $J$ and $P$ are the
traces of
\begin{equation}
J_{ab}=\frac{1}{3}%
(2KK_{ac}K_{b}^{c}+K_{cd}K^{cd}K_{ab}-2K_{ac}K^{cd}K_{db}-K^{2}K_{ab}),
\label{Jab}
\end{equation}%
and
\begin{eqnarray}
P_{ab} &=&\frac{1}{5}%
\{[K^{4}-6K^{2}K^{cd}K_{cd}+8KK_{cd}K_{e}^{d}K^{ec}-6K_{cd}K^{de}K_{ef}K^{fc}+3(K_{cd}K^{cd})^{2}]K_{ab}
\nonumber \\
&&-(4K^{3}-12KK_{ed}K^{ed}+8K_{de}K_{f}^{e}K^{fd})K_{ac}K_{b}^{c}-24KK_{ac}K^{cd}K_{de}K_{b}^{e}
\nonumber \\
&&+12(K^{2}-K_{ef}K^{ef})K_{ac}K^{cd}K_{db}+24K_{ac}K^{cd}K_{de}K^{ef}K_{bf}%
\}.  \label{Pab}
\end{eqnarray}

In general the action $\mathcal{I}_{G}$, the Hamiltonian and other
associated conserved quantities diverge when evaluated on the
solutions. Due to the fact that our spacetime is asymptotically
adS, one can use the systematic method to regulate the
gravitational action of asymptotically adS solutions which is
through the use of the counterterm method. It was shown that the
counterterm approach become quite reasonable when applied to
AdS/CFT, as the boundary counterterm has a natural interpretation
as conventional field theory counterterm that show up in the dual
CFT \cite{Mal}.

The counterterm action is a functional of the boundary curvature
invariants and do not affect on the symmetries and field equations
of the bulk $\mathcal{M}$
\begin{equation}
{I_{\mathrm{ct}}}=\int_{\partial \mathcal{M}}{{d^{n}}x\sqrt{-h}L(l,\widehat{R%
},\nabla \widehat{R},...)}.  \label{Ict1}
\end{equation}

In a general manner, the counterterm in Lovelock gravity is a
scalar constructed from the curvature invariants of the boundary
as in the case of Einstein gravity \cite{Kraus,DehNew}. Although
one can use the procedure of Ref. \cite{DehNew} to compute the
counterterm action for arbitrary horizon topology, for the sake of
brevity and simplification, we deal with the spacetime with zero
curvature boundary ($\widehat{R}_{abcd}(\gamma )=0$). In this case
all the counterterm containing the curvature invariants of the
boundary are zero (see \cite{TOLNLED,DehShah2} for more details)
and the counterterm reduces to
\begin{equation}
\mathcal{I}_{\mathrm{ct}}=\frac{1}{8\pi }\int_{\partial
\mathcal{M}}d^{n}x\sqrt{-\gamma }\left(
\frac{n-1}{l_{\mathrm{eff}}}\right) ,\   \label{Ict}
\end{equation}%
where $l_{\mathrm{eff}}$ is given by

\begin{equation}
l_{\mathrm{eff}}=\frac{15\sqrt{\alpha \left[ 1-\left( 1-\frac{3\alpha }{l^{2}%
}\right) ^{1/3}\right] }}{9\left( 1+\frac{\alpha }{l^{2}}\right) -\left[
2+\left( 1-\frac{3\alpha }{l^{2}}\right) ^{1/3}\right] ^{2}},  \label{Leff}
\end{equation}
It is notable that the effective $l_{\mathrm{eff}}$ reduces to $l$\ as $%
\alpha $\ goes to zero. Having the finite action and using the
Brown--York method of a quasilocal definition \cite{BY} with Eq.
(\ref{Ibulk})-(\ref{Ict}), one can introduce a divergence-free
stress-energy tensor as follows
\begin{equation}
T^{ab}=\frac{2}{{\sqrt{-\gamma }}}\frac{{\partial (\mathcal{I}_{G}+\mathcal{I%
}_{\mathrm{ct}})}}{{\partial \gamma {_{ab}}}}=\frac{1}{8\pi }\left[
(K^{ab}-K\gamma ^{ab})+2\alpha (3J^{ab}-J\gamma ^{ab})+\frac{n-1}{l_{\mathrm{%
eff}}}\gamma ^{ab}\right] .  \label{Tab}
\end{equation}%
The quasilocal conserved quantities associated with the
stress-energy tensor of Eq. (\ref{Tab}) can be defined as
\begin{equation}
Q(\xi )=\int_{\mathcal{B}}d^{n-1}\varphi \sqrt{\gamma }T_{ab}n^{a}\xi ^{b},
\label{Conserved}
\end{equation}%
where the the timelike unit vector $n^{a}$ is normal to the boundary $B$ and
$\xi ^{b}$ is the Killing vector. Regarding temporal Killing vector $\xi
=\partial /\partial t$ and taking into account Eqs. (\ref{Conserved}) and (%
\ref{Tab}), we can calculate the mass per unit volume $V_{n-1}$ as
\begin{equation}
M=\frac{\left( n-1\right) }{16\pi }m.  \label{Massk0}
\end{equation}%
We should note that the parameter $m$ can be calculated by using of the fact
that the metric function vanishes at the event horizon, $r_{+}$. Although
one can check that the form of Eq. (\ref{Massk0}) is valid for $k=\pm 1,0$,
we should indicate that, unlike $k=\pm 1$ cases, the mass parameter, $m$,
does not depend on the Lovelock parameter for the boundary flat solutions.

Although we used the counterterm method to calculate the finite
mass, one may find different methods in the literature for
computing the finite mass. It will be interesting to study the
conditions that enable those prescriptions to provide the right
mass for the solutions obtained here.

One of the best known prescriptions is that of
Arnowitt-Deser-Misner (ADM), which can be most applied in
asymptotically flat spacetimes. In addition, the ADM method may
also be applied to asymptotically anti-de Sitter space \cite{ADM}.
In such case, the mass may be extracted by comparison to a
suitable reference background (e.g. vacuum adS). Furthermore, we
refer the reader to the Ashtekar-Magnon-Das (AMD) formula
\cite{AMDmass}, the Hamiltonian method of Regge and Teitelboim
\cite{RTmass}, the generalized Komar integral of Lovelock gravity
\cite{Komar} and the subtraction method of Brown and York
\cite{BY,BY2}.

\end{document}